\documentclass[prl,aps,showpacs,twocolumn]{revtex4}
\usepackage{epsfig}
\usepackage{bm}

\begin{document}

\title{Intermittency and the passive nature of the magnitude of the magnetic field}

\author{\small A. Bershadskii$^{1,2}$ and K.R. Sreenivasan$^2$}
\affiliation{\small {\it $^1$ICAR, P.O. Box 31155, Jerusalem 91000, Israel}\\
{\it $^2$International Center for Theoretical Physics, Strada
Costiera 11, I-34100 Trieste, Italy}}

\begin{abstract}
It is shown that the statistical properties of the {\it magnitude}
of the magnetic field in turbulent electrically conducting media
resemble, in the inertial range, those of passive scalars in fully
developed three-dimensional fluid turbulence. This conclusion,
suggested by the data from Advanced Composition Explorer, is
supported by a brief analysis of the appropriate
magnetohydrodynamic equations.
\end{abstract}

\pacs{47.27.-i, 47.65.+a, 47.27.Qb, 47.27.Jv}

\maketitle

In recent years, the problem of turbulent advection and diffusion
of scalar and vector fields, both passive and active, has received
renewed attention (see, for instance, Refs.\
\cite{shr}-\cite{ching} and the papers cited there). Corresponding
experimental results are still few in number, partly because one
is still grappling with the precise correspondence between the
hypotheses of the theory and their experimental realization. We
discuss here one of the most interesting examples considered
recently: the case of magnetohydrodynamics (MHD) in which the
magnetic field fluctuation ${\bf B}$ is described by the equation
$$
\frac{\partial {\bf B}}{\partial t} = \nabla \times ({\bf v}
\times {\bf B}) + \eta \nabla^2 {\bf B}.    \eqno{(1)}
$$
Here, ${\bf v}$ is the turbulence velocity and the $\eta$ is
magnetic diffusivity. Equation (1) can be regarded as a vector
analogue of the advection-diffusion equation
$$
\frac{\partial \theta }{\partial t}=-({\bf v} \cdot \nabla) \theta
+ D \nabla^2 \theta              \eqno{(2)}
$$
for the evolution of a passive scalar $\theta$ subject to
molecular diffusivity $D$. Aside from the fact that ${\bf B}$ is a
vector and $\theta$ a scalar, the equations are different also
because $\bf v$ in Eq.\ (1) can be affected quite readily by the
feedback of the magnetic field $\bf B$. Our interest here is to
explore the extent of similarities, despite these obvious
differences, in the inertial range statistics of the magnetic and
passive scalar fields.

\begin{figure} \vspace{-1.5cm}\centering
\epsfig{width=.45\textwidth,file=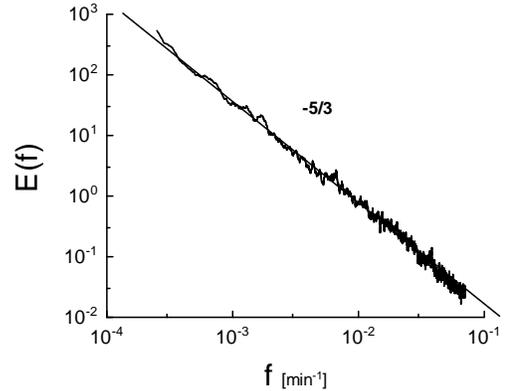} \vspace{-4cm}
\caption{Energy spectrum of the {\it magnitude} $B = \sqrt{B_i^2}$
of the magnetic field $\bf B$ in the solar wind plasma, as
measured by the ACE magnetometers in the nanoTesla range for the
year 1998 (4 min average).}
\end{figure}

Solar wind is an excellent natural ``laboratory" for the MHD
problem. It is known that the statistical properties of velocity
fluctuations in the solar wind are remarkably similar to those
observed in fluid turbulence \cite{bur}. It is also known that the
plasma power spectra of the magnetic field and velocity
fluctuations often contain an ``inertial'' range with a slope of
approximately $-5/3$ (for reviews, see \cite{gold},\cite{bur}).
The approximately $-5/3$ power-law is especially common for
magnitude fluctuations $B = \sqrt{B_i^2}$ (the summation over
repeated indexes is assumed) of the magnetic field, as one can see
in Fig.\ 1. For computing this spectrum, we have used the data
obtained from Advanced Composition Explorer (ACE) satellite
magnetometers for the year 1998. In this period, the sun was quiet
and the data are statistically stable. The range of scales for
which the ``$-5/3$" power holds is taken to be the inertial range;
the smaller scales are obliterated because of the instrument
resolution and the truncation at the large-scale end is governed
by the record length chosen for Fourier transforming.

The nature of the spectrum for each individual component of the
magnetic field is more variable from one component of $\bf B$ to
another, and from one situation to another, perhaps because of
large anisotropies in the magnetic field $\bf B$, but the result
for the $magnitude$ of $\bf B$ seems more robust. Scaling spectrum
with the $-5/3$ slope (Fig. 1) is quite typical of that observed
for passive scalar fluctuations in fully developed
three-dimensional fluid turbulence (the so-called Corrsin-Obukhov
spectrum \cite{my}). Spurred by this similarity, we were motivated
to explore further the properties of the magnitude $B$, and
compare them with those of the passive scalar. More detailed
statistical information is provided by the structure functions
scaling
$$
\langle |\Delta B_{\tau}|^p \rangle \sim \tau^{\zeta_p},
\eqno{(3)}
$$
where
$$
\Delta B_{\tau} =B(t+\tau)-B(t).       \eqno{(4)}
$$
The exponent $\zeta_2$ is directly related to the spectral
exponent (in our case $\zeta_2 \approx 5/3 - 1 = 2/3$ \cite{my}).
If the dependence of $\zeta_p$ on $p$ is nonlinear, it is
well-known that one has to deal with intermittency.

\begin{figure}
\centering \epsfig{width=.45\textwidth,file=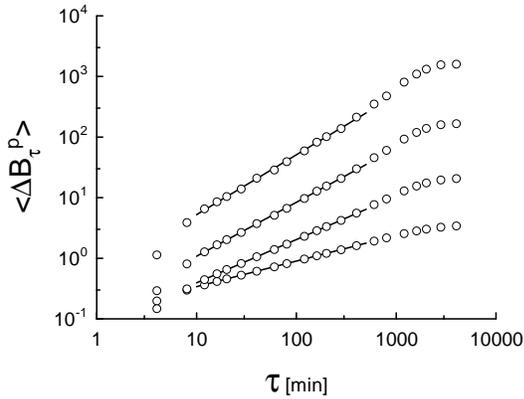}
\vspace{-4cm} \caption{Structure functions of magnetic field {\it
magnitude} in the solar wind plasma as measured by the ACE
magnetometers in nanoTesla for the year 1998 (4 min averages). The
straight lines (the best fits) are drawn to indicate the scaling
law (3) in the inertial range.}
\end{figure}
\begin{figure}
\centering \epsfig{width=.45\textwidth,file=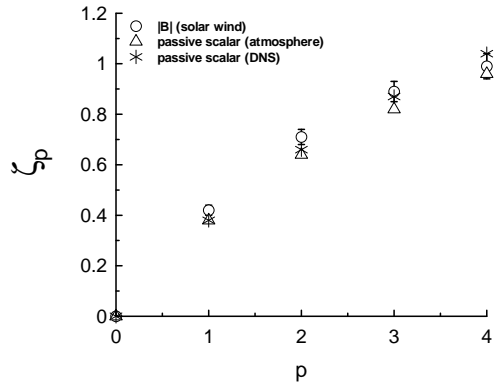}
\vspace{-4cm} \caption{Scaling exponents (3) calculated for $B$ in
the solar wind (circles) and for the passive scalar in the
atmospheric turbulence (triangles, \cite{schmidt}), and in the
direct numerical simulation of 3D fluid turbulence (stars,
\cite{gotoh})}
\end{figure}
\begin{figure}
\centering \epsfig{width=.45\textwidth,file=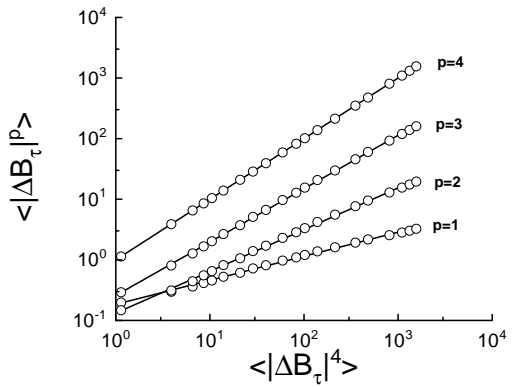}
\vspace{-4cm} \caption{Extended self-similarity (ESS) of the
magnetic field {\it magnitude} in the solar wind plasma. The
straight lines (the best fit) are drawn to indicate the ESS (6).}
\end{figure}

\begin{figure}
\centering \epsfig{width=.45\textwidth,file=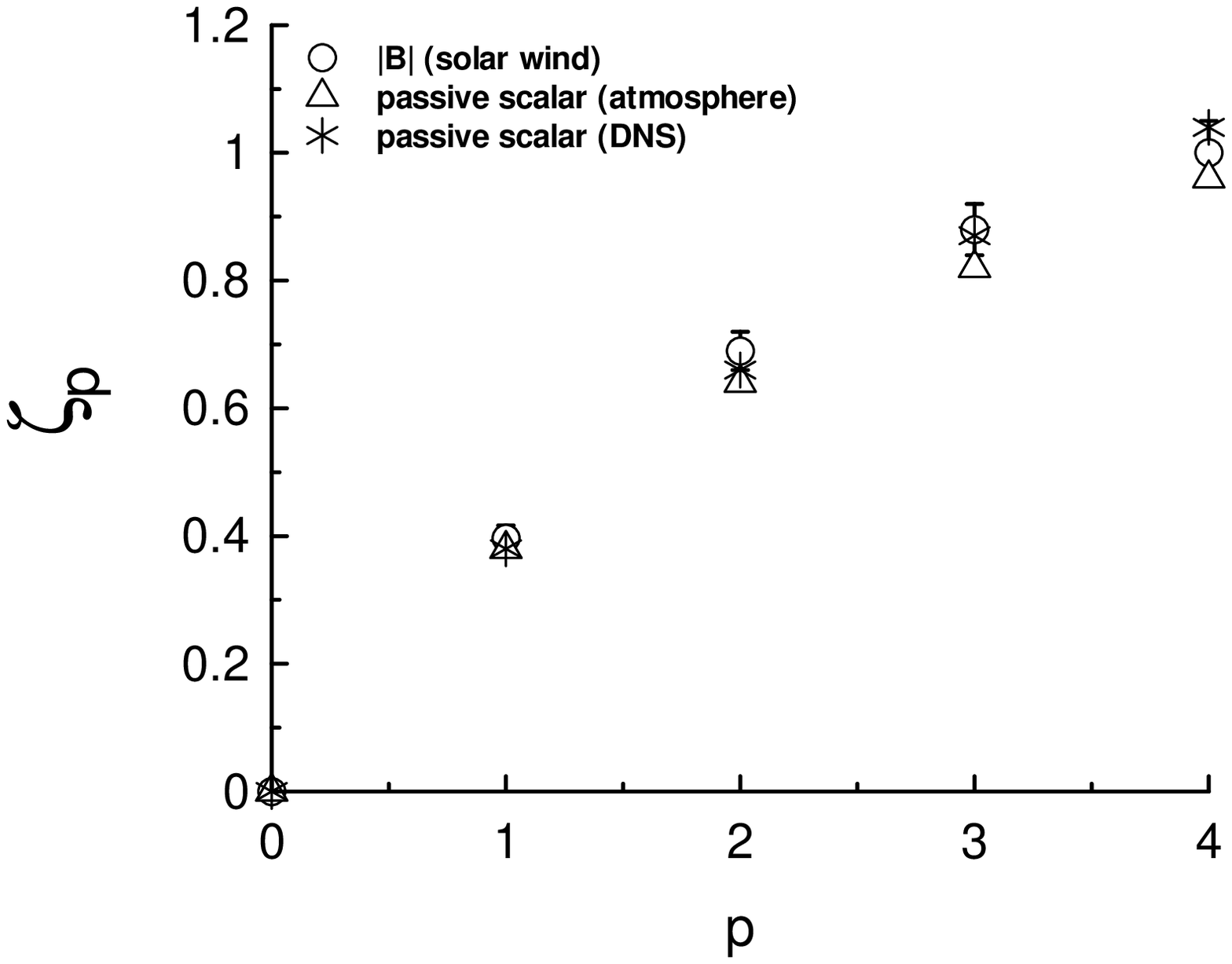}
\vspace{-4cm} \caption{The same as in Fig.\ 3 but using the ESS
method (6) for $B$.}
\end{figure}
Figure 2 shows the scaling of structure functions for the solar
wind data. Slopes of the straight-line fits in the apparently
scaling region provide us the scaling exponents $\zeta_p$; these
are shown in Fig.\ 3 as circles. Triangles in the figure indicate
experimental values obtained for temperature fluctuations in the
atmosphere \cite{schmidt}. The other experimental data
\cite{other1},\cite{other2} are in agreement with each other to
better than 5$\%$. The $\star$ symbols are for the passive scalar
field obtained by numerically solving the advection-diffusion in
three-dimensional turbulence \cite{gotoh}. It is clear that the
exponents for the passive scalar data are in essential agreement
with those for the {\it magnitude} fluctuations of the magnetic
field.

One can, in fact, analyze the solar wind data somewhat differently
using the notion of the extended self-similarity (ESS). Since,
empirically, the fourth order exponent is quite closely equal to 1
for the magnetic field magnitude, i.e.,
$$
\langle |\Delta B_{\tau}|^4\rangle \sim \tau, \eqno{(5)}
$$
we can extend the scaling range (and consequently improve the
confidence with which those exponents are determined) by
redefining them as
$$
\langle |\Delta B_{\tau}|^p \rangle \sim \langle |\Delta
B_{\tau}|^4\rangle^{\zeta_p}. \eqno{(6)}
$$

Figure 4 shows the ESS dependence (6). The slopes of the best-fit
straight lines in this figure provide us with the ESS scaling
exponents $\zeta_p$, which are shown in Fig.\ 5 as circles. The
other symbols have remained unchanged from Fig.\ 3. The shift of
the exponents $\zeta_p$ in comparison to those from ordinary
self-similarity is about 4$\%$, but the scaling interval for ESS
is considerably larger. This increased scaling range is well-known
in other contexts \cite{ben}.

The results shown in Figs.\ 3 and 5 suggest that at least up to
the level of the fourth-order the scaling exponents for the
passive scalars and for the magnitude of the magnetic field are
essentially the same. This is both surprising and
thought-provoking, and needs to be understood further. To this
end, let us return to Eq.\ (1) and specialize \cite{comment}, for
simplicity, to the case of incompressible moving medium ($\nabla
\cdot{\bf v} =0$). Equation (1) can then be rewritten as
$$
\frac{\partial {\bf B}}{\partial t} =- ({\bf v} \cdot \nabla) {\bf
B} + ({\bf B} \cdot \nabla) {\bf v}  +\eta \nabla^2 {\bf B}.
\eqno{(7)}
$$
Let us now consider the equation for the magnitude $B$ of the
magnetic fluctuations given by ${\bf B} = B {\bf n}$, where ${\bf
n}$ is the unit vector with its direction along ${\bf B}$:
$n_i=B_i/B$. Multiplying both sides of Eq.\ (7) by the vector
${\bf n}$ and taking into account that $n_i^2 =1$ we obtain
$$
\frac{\partial B}{\partial t}= -({\bf v}\cdot \nabla) B + \eta
\nabla^2 B + \lambda B, \eqno{(8)}
$$
in which the ``friction-stretching" (or the production)
coefficient $\lambda$ in the last term has the form
$$
\lambda = n_in_j \frac{\partial v_i}{\partial x_j} - \eta \left(
\frac{\partial n_i}{\partial x_j}\right)^2,  \eqno{(9)}
$$
with the indices $i$ and $j$ representing the space coordinates,
and the summation over repeated indexes is assumed. The first term
on the right hand side of Eq.\ (9) is crucial for any dynamo
effect.

If the statistical behaviors of $\theta$ and $B$ are to be
similar, as suggested by Figs.\ 1, 3 and 5, we should be able to
observe the underlying similarity between Eqs.\ (2) and (8). There
is a major difference corresponding the presence in Eq.\ (8) of
the production term $\lambda B$. However, given the empirical
indications that $B$ and $\theta$ are similar in the inertial
range, it is appropriate to look for circumstances under which the
$\lambda$ term in Eq.\ (8) may be small. The second term in
$\lambda$ is assured to be small because the smallness of the
magnetic diffusivity $\eta$, but difficulties may arise from the
first term on the right hand side of Eq.\ (9).

To eliminate {\it directional} dependencies in Eq.\ (8), let us
make the following conditional average of that equation. That is,
fix the magnitude $B$ in the vector field ${\bf B}=B{\bf n}$ while
performing the average over all realizations of the direction
vector field ${\bf n}$ permitted by the vector equation (7). Let
us denote this ensemble average as $\langle ... \rangle_{{\bf
n}}$. From the definition, this averaging procedure does not
affect $B$ itself, but affects the velocity field ${\bf v}$ and
the ``friction-stretching" coefficient $\lambda$ in Eq.\ (8). We
thus obtain
$$
\frac{\partial B}{\partial t}= -(\langle {\bf v} \rangle_{{\bf n}}
\cdot \nabla) B + \eta \nabla^2 B + \langle \lambda \rangle_{{\bf
n}}B. \eqno{(10)}
$$
It is worth emphasizing that the solutions of the original
equation (7) satisfy Eqs.\ (8) and (10), but not all possible
formal solutions of the Eqs.\ (8) and (10) satisfy Eq.\ (7);
similarly, not all formal solutions of Eq.\ (10) satisfy Eq.\ (8)
while all solutions of Eq.\ (8) do satisfy Eq.\ (10). Restricting
comments to the relationship between Eqs.\ (8) and (10), the
solutions of the two equations are the same only if the initial
conditions are the same and if realizations of $\langle {\bf v}
\rangle_{\bf n}$ and of $\langle \lambda \rangle_{\bf n}$, related
to these initial conditions by the conditional average procedure,
are obtained from solutions applicable to Eq.\ (8).

Returning now to Eq.\ (10), the conditionally averaged velocity
field $\langle {\bf v} \rangle_{{\bf n}}$ may posses statistical
properties that are different from those of the original velocity
field ${\bf v}$, and there can be circumstances under which
$\langle \lambda \rangle_{{\bf n}}=0$, or small. If so, the
similarity between Eqs.\ (2) and (10) (and, consequently, Eq.\
(8)) can be the basis for the similarity in statistical properties
of their solutions. Therefore, finding conditions under which
$\langle \lambda \rangle_{{\bf n}}=0$, or small, seems to be a
useful exercise.

It is, however, difficult to guess {\it a~prioiri} when $\langle
\lambda \rangle_{{\bf n}}$ is negligible, because there is no
small parameter for the stretching part of $\lambda$. Therefore,
let us consider a generic set of conditions, presumably for the
inertial range, which can result in $\langle n_in_j
\partial v_i/\partial x_j \rangle_{{\bf n}} =0$. This can be a
combination of isotropy, which yields
$$
\langle n_in_j\rangle_{{\bf n}} =0 ~~~~(i\neq j)
$$
and
$$
 \langle n_1^2\rangle_{{\bf n}}=\langle n_2^2\rangle_{{\bf n}}=
\langle n_3^2 \rangle_{{\bf n}}, \eqno{(11)}
$$
and statistical independence
$$
\langle n_in_j \varphi \rangle_{{\bf n}} = \langle n_in_j
\rangle_{{\bf n}}\langle  \varphi \rangle_{{\bf n}}, \eqno{(12)}
$$
where $\varphi = \partial v_k/\partial x_l$ for arbitrary $k$ and
$l$.

We should emphasize that the conditional average indicated by
$\langle \dots \rangle_{{\bf n}}$ and the global average indicated
by $\langle \dots \rangle$ are quite different; because of this,
the quantity $B$ in (10) remains a fluctuating variable. To
eliminate the stretching part from the conditionally averaged
coefficient $\langle \lambda \rangle_{{\bf n}}$---this being
critical for explaining the observed similarity in the scaling of
structure functions between $B$ and $\theta$---one does not need
to satisfy conditions (11) and (12) for all realizations of the
magnetic field ${\bf B}$, but only for the subset of realizations
that gives the main statistical contribution to the structure
functions (3). Let us name this subset of realizations as {\it I}.
The structure functions (3) depend on the statistical properties
of the {\it increments} with respect to $\tau$, namely $\Delta
B_{\tau}$, belonging to the inertial range of scales. One of the
consequences of intermittency is that the statistical properties
of the increments are essentially different from those of the
field ${\bf B}$ itself. Therefore, the subset {\it I} need not
generally coincide with the subset {\it G}, say, that gives the
main statistical contribution to the {\it global} average $\langle
n_in_j \partial v_k/\partial x_l \rangle$. This means, in
particular, that the conditions (11) and (12) can be valid for the
inertial interval (i.e. for subset {\it I}), while globally (i.e.
for subset {\it G}) these conditions could well be violated.

We now use conditions (11) and (12) in the presence of the
incompressibility condition $\partial v_i/\partial x_i=0$ and
obtain
$$
\langle \lambda \rangle_{{\bf n}} = -\eta \langle \left(
\frac{\partial n_i}{\partial x_j}\right)^2 \rangle_{{\bf n}}.
\eqno{(13)}
$$
That is, the difference between the passive scalar equation (2)
and the conditionally averaged equation (10) for $B$ is reduced to
pure ``friction" with the friction coefficient given by (13).
Equation (10) can then be reduced in Lagrangian variables to
$$
\frac{dB}{dt}=\langle \lambda \rangle_{{\bf n}} B, \eqno{(14)}
$$
with the ``multiplicative noise" $\langle \lambda \rangle_{{\bf
n}}$ given by Eq.\ (13). Weak diffusion of Lagrangian ``particles"
can be described as their wandering around the deterministic
trajectories. Introduction of a weak diffusion is equivalent to
introduction of additional averaging in Eq.\ (14) over random
trajectories \cite{zeld1}. The small parameter $\eta$ in (13) and
(14) will then determine a slow time in comparison with the time
scales in the inertial interval and will therefore not affect
scaling properties of $B$ in the inertial interval. This explains
the similarity of scaling between $B$ and $\theta$.

In summary, we have shown that remarkable scaling similarities
exist {\it in the inertial range} between the passive scalar and
the {\it magnitude} of the magnetic field in MHD flows. Motivated
by this observation, we have derived the dynamical equation for
$B$ and argued that, under circumstances governed by Eqs.\ (11)
and (12), dynamical similarity exists between the equations
governing the passive scalar and the {\it magnitude} of the
magnetic fluctuations. The conditions under which (11) and (12)
are valid, perhaps best described as ``directional randomness",
may be quite general and applicable to other circumstances
(though, perhaps, not for the vorticity equation whose form is
strongly nonlinear in any turbulent situation).

We are grateful to ACE/MAG instrument team as well as to the ACE
Science Center for providing the data and support, to T. Gotoh for
providing Ref. \cite{gotoh} before publication, and to G.
Falkovich, to J. Schumacher and to V. Yakhot for comments.

\end{document}